\documentclass[aps,prl,twocolumn,superscriptaddress,showpacs]{revtex4}
\usepackage{graphicx}
\begin{document}

\title{Break--Junction Tunneling on MgB$_2$}

\author{H.\ Schmidt}
\affiliation{Materials Science Division, Argonne National
Laboratory, Argonne, IL 60439, USA} \affiliation{Physics
Division,Illinois Institute of Technology, Chicago, IL 60616, USA}

\author{J.\ F.\ Zasadzinski}
\affiliation{Materials Science Division, Argonne National
Laboratory, Argonne, IL 60439, USA} \affiliation{Physics
Division,Illinois Institute of Technology, Chicago, IL 60616, USA}

\author{K.\ E.\ Gray}
\affiliation{Materials Science Division, Argonne National
Laboratory, Argonne, IL 60439, USA}

\author{D.\ G.\ Hinks}
\affiliation{Materials Science Division, Argonne National
Laboratory, Argonne, IL 60439, USA}

\date{\today}

\begin{abstract}
Tunneling data on magnesium diboride, MgB$_2$, are reviewed with a
particular focus on superconductor--insulator--superconductor
(SIS) junctions formed by a break--junction method. The collective
tunneling literature reveals two distinct energy scales, a large
gap, $\Delta_L\sim7.2$~meV, close to the expected BCS value, and a
small gap, $\Delta_S\sim2.4$~meV. The SIS break junctions show
clearly that the small gap closes near the bulk critical
temperature, $T_c=39$~K. The SIS spectra allow proximity effects
to be ruled out as the cause for the small gap and therefore make
a strong case that MgB$_2$ is a coupled, two--band superconductor.
While the break junctions sometimes reveal parallel contributions
to the conductance from both bands, it is more often found that
$\Delta_S$ dominates the spectra. In these cases, a subtle feature
is observed near $\Delta_S+\Delta_L$ that is reminiscent of
strong--coupling effects. This feature is consistent with
quasiparticle scattering contributions to the interband coupling
which provides an important insight into the nature of two--band
superconductivity in MgB$_2$.
\end{abstract}


\maketitle

\section{Introduction}
Approximately one and a half years after the discovery
\cite{jNagamatsu01} of superconductivity in MgB$_2$ a wealth of
information about its properties has been collected. While its
high critical temperature and simple crystal structure generated
an initial flurry of activity, there is now an increasing interest
in the nature of superconductivity in MgB$_2$, as it appears to be
one of the rare examples of a two--band superconductor.

Theoretically, such two--band behavior was  proposed for MgB$_2$
because the electronic system consists of two qualitatively
different types of charge carriers, derived from boron $\pi$-- and
$\sigma$--bands, respectively \cite{ayLiu01}. The $\pi$--bands are
three dimensional (3D), while the $\sigma$--bands are effectively
restricted to two dimensions (2D). For convenience, these actual
four bands are henceforward treated as two effective bands.
Superconductivity is proposed to arise from electron--phonon
coupling of the 2D band with a specific boron bond--stretching
mode. That implies, that superconductivity originates from the
$\sigma$--band and that superconductivity in the $\pi$--band is
driven by that primary interaction. As a result, the properties of
superconductivity are proposed to be different in both bands,
provided the material is in the clean limit and the electronic
systems do not strongly mix. The 2D band shows a large gap,
$\Delta_L\sim7.2$~meV, whereas the 3D band shows a small gap,
$\Delta_S\sim2.4$~meV, both closing at a joint critical
temperature, $T_c$. In the dirty limit, only one gap of
intermediate magnitude is expected to be observed, closing at a
reduced $T_c$.

There now is sufficient experimental evidence to strongly support
such a two--band scenario, and there is a growing consensus in the
community to this end. However, a large fraction of these
experiments involved tunneling spectroscopy which is susceptible
to surface imperfections. Specifically, surface proximity effects
can mimic the effects of two--band superconductivity. In this
paper, we will present an overview of tunneling results, and will
use this together with our own experimental data to argue in favor
of the two--band nature of superconductivity in MgB$_2$.

\begin{figure*}
\vskip-0.1in
\includegraphics[scale=0.81]{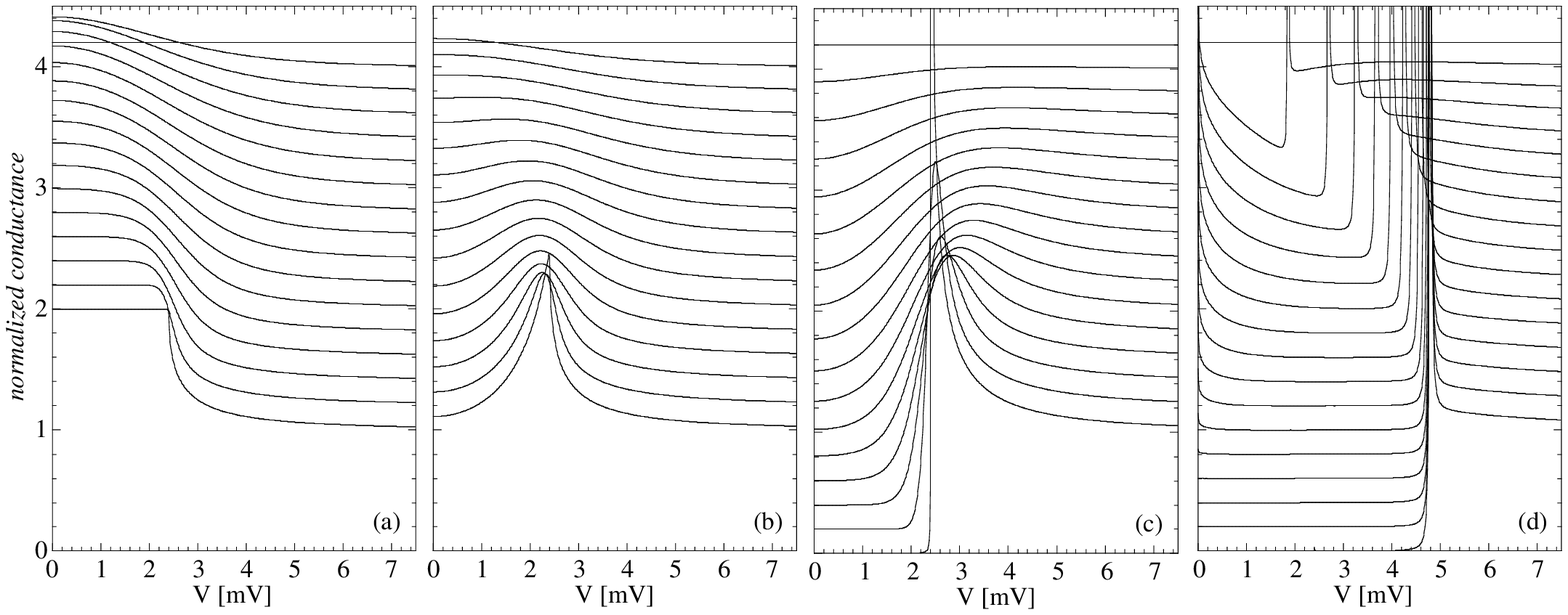}
\vskip-0.1in
\caption{\label{temp}{Simulated temperature dependence
of (a) an SIN junction with $Z=0$ (purely metallic), (b) $Z=0.5$
(intermediate), (c) $Z\rightarrow\infty$ (pure tunneling) and (d)
an SIS tunnel junction. Parameters used are: $\Delta_0=2.4$~meV
and $T_c={\Delta_0\over1.76k_B}=15.8$~K. Curves are vertically
shifted for clarity, temperature increment is 1~K. To avoid
singularities, $\Gamma=1$~$\mu$eV was included (using the
formalism described in Ref.\ \cite{aPlecenik94} for (a) and
(b)).}}
\end{figure*}

\section{Techniques of Tunneling Spectroscopy}

Tunneling spectroscopy traditionally presents one of the most
direct probes of the superconducting energy gap \cite{iGiaver60}.
In superconductor--insulator--normal metal (SIN) tunnel junctions,
the conductance at low temperature is equivalent to the density of
states (DOS) near the Fermi energy, $E_F$. At finite temperature,
the conductance may be calculated from:
\begin{equation}
{dI\over dV}=-\sigma_N\int N(E){\partial f(E+eV)\over\partial
eV}dE,
\end{equation}
where $N$ is the superconducting DOS and $\sigma_N$ the
conductance of the junction in the normal state. To account for
experimentally observed broadening in excess of thermal smearing,
a smearing parameter $\Gamma$ usually is introduced into the BCS
DOS \cite{rcDynes78}:
\begin{equation}
N(E)=\Re{|E|-i\Gamma\over\sqrt{(|E|-i\Gamma)^2-\Delta^2}}.\label{DOS}
\end{equation}
A widely used realization of such SIN junctions is the scanning
tunneling microscope (STM), where the barrier is given by a short
length of vacuum that separates the STM tip from the sample. This
technique combines an ideal insulating barrier with high spatial
resolution. On the other hand, quite frequently a tip is placed
into direct contact to the sample, a setup generally known as
point contacts. Such systems usually lack fine spatial resolution,
but they allow for a significantly lower contact resistance and
can be used to realize a continuous transition from a tunnel
junction (insulating barrier) to a metallic contact (no barrier).
Metallic contacts show enhanced current at bias voltages lower
than the superconducting energy gap due to Andreev reflections
\cite{afAndreev64}, and hence they also allow one to determine the
energy gap. In the intermediate case, both tunneling and Andreev
reflections contribute to the transport over the barrier, a case
that was analytically described by Blonder, Tinkham and Klapwijk
(BTK) \cite{geBlonder82}.

Finally, symmetrical SIS junctions display very sharp features at
twice the energy gap, that are a result of the convolution of two
BCS DOSs:
\begin{equation}
I={\sigma_N\over e}\int N(E)N(E+eV)[f(E)-f(E+eV)]dE.\label{SIS}
\end{equation}
Such SIS junctions may be used to study the energy gap and a major
advantage is their ability to trace the gap feature to highest
temperatures, as the coherence peaks prove to be fairly
insensitive against thermal smearing.

To demonstrate this, Fig.\ \ref{temp} compares the temperature
evolution of simulated spectra assuming a 2.4~meV BCS
superconductor with insignificant smearing ($\Gamma=1$~$\mu$eV),
for (a) a purely metallic, (b) an intermediate and (c) a pure
tunnel junction, as well as (d) a symmetric SIS tunnel junction.
Not only do the coherence peaks remain sharp up to highest
temperatures in the SIS case, their position furthermore directly
scales with the closing of the gap. In comparison, the peak
position in the SIN tunneling case increases as $T$ approaches
$T_c$---even though the gap actually closes. A close relation
between the SIS coherence peak position and $\Delta(T)$ is
preserved even in the presence of significant smearing. We will
make use of this peculiarity to precisely measure $\Delta_S(T)$.

\begin{table*}
\begin{tabular}{p{0.2\textwidth}p{0.13\textwidth}p{0.18\textwidth}
p{0.17\textwidth}p{0.17\textwidth}p{0.1\textwidth}}
\hline\hline \multicolumn{3}{l}{{Experiment}}&
\multicolumn{2}{l}{\hskip0.4cm{Energy Gap / [meV]}}&{Reference}\\
\hline
{STM} &&&{$\Delta_S=2.0$}&&
{\cite{gRubio-Bollinger01,hSuderow02}}\\
&&&\multicolumn{2}{c}{\hskip-0.5in{$\Delta=5-7$}}&
{\cite{aSharoni01a,aSharoni01b}}\\
&&&{$\Delta_S=2.3$}&{$\Delta_L=7.1$}&
{\cite{gKarapetrov01,gKarapetrov01b,Iavarone-unpublished,Olsson-unpublished}}\\
&&&{$\Delta_S=3.5$}&{$\Delta_L=7.5$}&
{\cite{fGiubileo01,fGiubileo02a,fGiubileo02b}}\\
&&&{$\Delta_S=2.2$}&{$\Delta_L=6.9$}&
{\cite{Karpinski-unpublished,Eskildsen-unpublished}}\\
&&&{$\Delta_{xy}$\hskip-0.02cm$=5.0$}&{$\Delta_z$\hskip0.15cm$=8.0$}&
{\cite{pSeneor01}}\\
\hline {Point Contact} &\multicolumn{2}{l}{{MgB$_2$--In, Ag
Paint}}&{$\Delta_S=2.6$}&&
{\cite{aPlecenik02}}\\
&\multicolumn{2}{l}{{MgB$_2$--Pt}}&{$\Delta_S=1.7$}&{$\Delta_L=7$}&
{\cite{fLaube01}}\\
&\multicolumn{2}{l}{{MgB$_2$--Nb}}&{$\Delta_S=3$}&{$\Delta_L=7$}&
{\cite{Dyachenko-unpublished}}\\
&\multicolumn{2}{l}{MgB$_2$--Nb,
PtIr}&{$\Delta_S=2.8$}&{$\Delta_L=9.8$}&
\cite{zzLi01,zzLi02b}\\
&\multicolumn{2}{l}{{MgB$_2$--Cu}}&{$\Delta_S=2.8$}&{$\Delta_L=7.0$}&
{\cite{pSzabo01}}\\
&\multicolumn{2}{l}{{MgB$_2$--Cu, Ag}}&{$\Delta_S=2.45$}&
{$\Delta_L=7.0$}&{\cite{ygNaidyuk02,Bobrov-unpublished,Yanson-unpublished}}\\
&\multicolumn{2}{l}{{MgB$_2$--Au}}&{$\Delta_S=2-3$}&{$\Delta_L=6-8$}&
{\cite{Lee-unpublished}}\\
&\multicolumn{2}{l}{{MgB$_2$--Au}}&{$\Delta_S=2.3$}&{$\Delta_L=6.2$}&
{\cite{yBugoslavsky02}}\\
&\multicolumn{2}{l}{{MgB$_2$--Au}}&\multicolumn{2}{c}{\hskip-0.5in{$\Delta=3-4$}}&
{\cite{aKohen01}}\\
&\multicolumn{2}{l}{{MgB$_2$--Au, Pb}}&{$\Delta_S=2.7$}&
{$\Delta_L=7.1$}&{\cite{Belogolovskii-private}}\\
&\multicolumn{2}{l}{{MgB$_2$--Au, Pt, PtIr, In, Ag
Paint}}&{$\Delta_S=2.8$}&{$\Delta_L=7.1$}&
{\cite{rsGonnelli02b,Gonnelli-unpublished,rsGonnelli02a}}\\
\hline {Break Junction} &&&{{$\Delta_S=1.7-2$}}&&
{\cite{rsGonnelli02a,rsGonnelli01}}\\
&&&{$\Delta_S=2.5$}&{$\Delta_L=7.6$}&
{\cite{hSchmidt01,hSchmidt01b,hSchmidt02,hSchmidt-unpublished}}\\
&&&$\Delta_S=2-2.25$&{{$\Delta_L=8.5-9.5$}}&
{\cite{Takasaki-unpublished}}\\
&&{(SQUID)}&{$\Delta_S=2.02$}&&
{\cite{yZhang01}}\\
&&&\multicolumn{2}{c}{\hskip-0.5in{N/A}}&
{\cite{yXuan01,zzLi02}}\\
\hline {Planar Junction} &{{Step--Edge}}&MgB$_2$/Ag/MgB$_2$
&\multicolumn{2}{c}{\hskip-0.5in{N/A}}&
{\cite{Kye-unpublished}}\\
&{{Ramp}}&MgB$_2$/MgO/MgB$_2$&\multicolumn{2}{c}{\hskip-0.5in{N/A}}&
{\cite{dMijatovic02,Mijatovic-unpublished}}\\
&{Nanobridge}&{(SQUID)}&\multicolumn{2}{c}{\hskip-0.5in{N/A}}&
{\cite{aBrinkman01b,Mijatovic-unpublished}}\\
&{Trench}&{(SQUID)}&\multicolumn{2}{c}{\hskip-0.5in{N/A}}&
{\cite{gBurnell01,gBurnell02,Burnell-unpublished}}\\
&\multicolumn{2}{l}{{Metal Masked Ion Damage}}
&\multicolumn{2}{c}{\hskip-0.5in{N/A}}&
{\cite{Kang-unpublished,Kang-unpublished2}}\\
\hline {Sandwich} &{MgB$_2$/Ag}&&&{$\Delta_L=7.3$}&
{\cite{dkAswal02}}\\
&{MgB$_2$/Pb}&&{$\Delta_S=1.75$}&
{$\Delta_L=8.2$}&{\cite{mhBadr02}}\\
&\multicolumn{2}{l}{{MgB$_2$/Al/Al$_2$O$_3$/Nb}}&
{$\Delta_S=2.2$}&&{\cite{gCarapella02}}\\
&\multicolumn{2}{l}{{MgB$_2$/AlN/NbN}}&{$\Delta_S=2.95$}&
&{\cite{Saito-unpublished}}\\
&\multicolumn{2}{l}{{MgB$_2$/MgO$_x$/Au, Ag}}&
{$\Delta_S=2.5$}&&{\cite{Ueda-unpublished}}\\
\hline\hline
\end{tabular}
\caption{\label{gaps}{Energy gaps of MgB$_2$ inferred from
different tunneling experiments in the literature. Experiments are
listed by technique as well as they are sorted by research group.
For each group, only the most recent findings are given. For
reference, all tunneling studies are listed, including a few that
did not determine any energy gap value.}}
\end{table*}

\section{Experiment}

A tremendous amount of tunneling work has been performed on
MgB$_2$, and even though such a narrow view certainly does not
mirror the full depth of information obtained from these studies,
we for now want to focus only on the inferred energy gap values.
Table \ref{gaps} gives a compilation of these values, and it
immediately becomes evident that most studies agree with the
presence of two gaps around $\Delta_S=2.0-2.8$~meV and
$\Delta_L=7.0-7.5$~meV. This observation of two gaps, as well as
their values, are in nice agreement with theoretical predictions
\cite{ayLiu01,hjChoi02a,hjChoi02b}, however, some more effort is
needed to positively confirm two--band superconductivity as the
origin of these features.

Tunneling spectroscopy by nature is a surface probe and it is
important to carefully verify that its results represent the bulk
properties of the sample. Degraded surfaces were repeatedly
suggested as an origin for small gap values \cite{hSchmidt01}, and
there still are some lingering concerns as to whether multiple
gaps in tunneling are {\lq\lq}real{\rq\rq}. To answer to that,
more information than the mere gap values is needed, and tunneling
can readily supply such information.

The polycrystalline samples used in the present work were prepared
as described in Ref.\ \cite{hSchmidt02}. Tunneling spectra were
taken using a home--built point--contact apparatus and a Au tip
\cite{lOzyuzer98}. This resulted in both SIN and SIS junctions,
however, henceforward we will focus on the SIS break junctions.

\begin{figure}[t]
\vskip-0.05in
\includegraphics[scale=0.45]{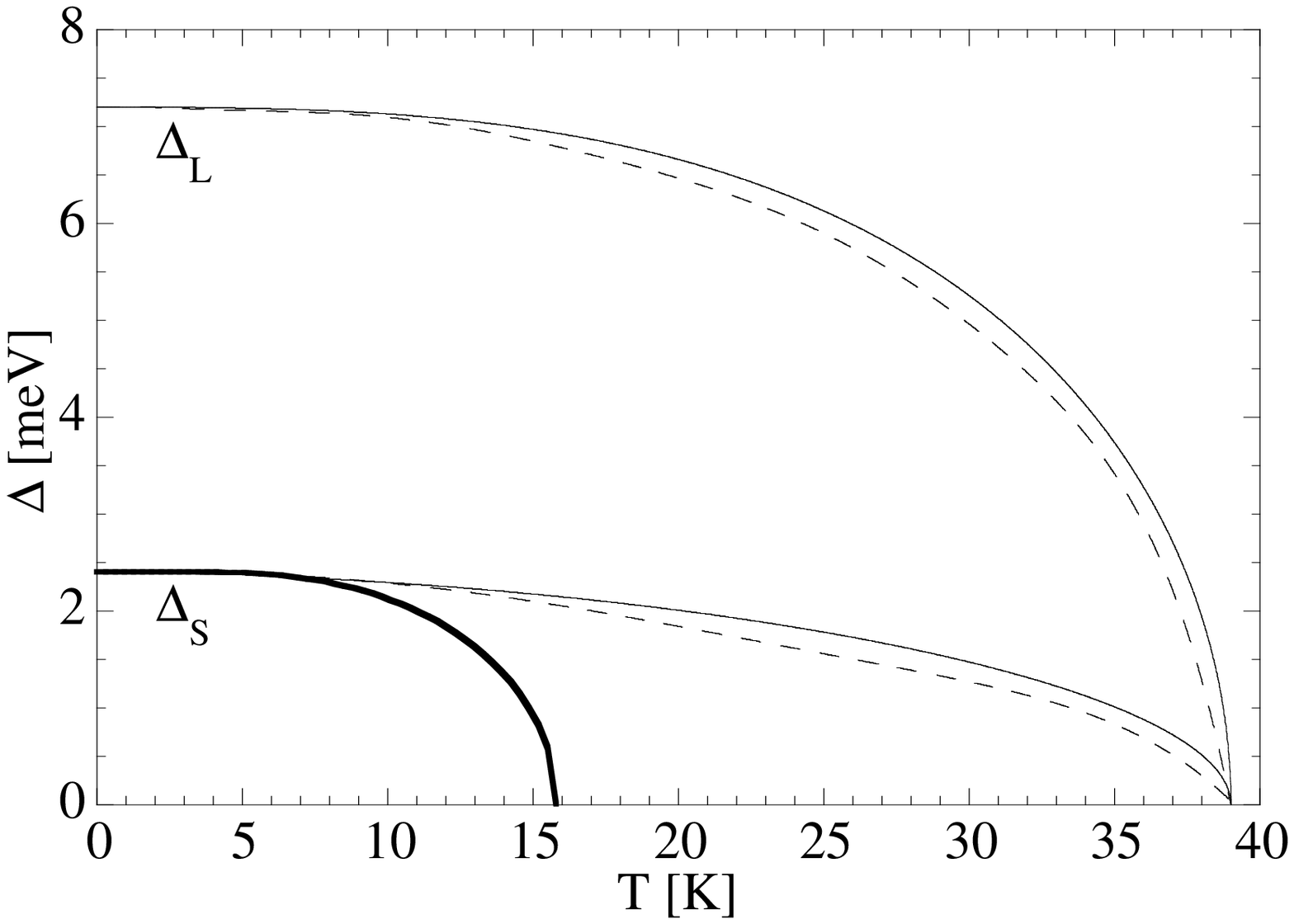}
\vskip-0.1in\caption{\label{DsT}{Temperature dependence for an
uncoupled small gap (heavy line, which was used to compute
Fig.~\ref{temp}) as compared to a small, coupled gap (thin line
after Ref.\ \cite{hjChoi02b}, dashed line after Ref.\
\cite{ayLiu01}).}}
\end{figure}

\begin{figure}[b]
\includegraphics[scale=0.45]{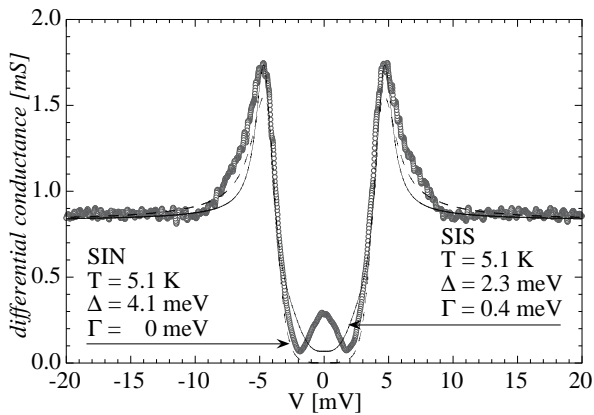}
\vskip-0.1in\caption{\label{sinsis}{Low temperature conductance
spectrum (symbols) taken from Ref.\ \cite{hSchmidt01} along with
fits to an SIN (dashed line) and an SIS model (solid line).}}
\end{figure}

\begin{figure}[t]
\includegraphics[scale=0.455]{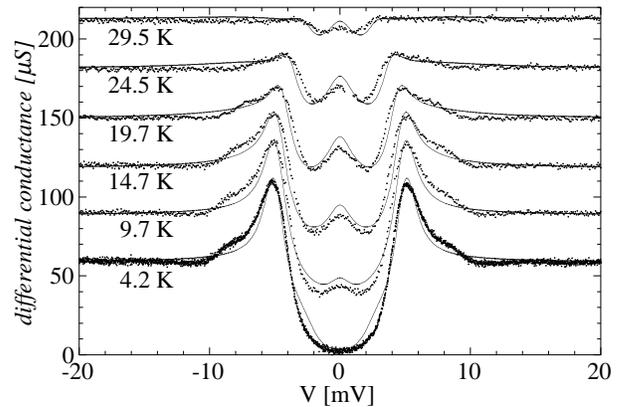}
\vskip-0.1in\caption{\label{temp-data}{Temperature dependence of
an SIS spectrum showing exclusively the small gap
\cite{hSchmidt02}.}}
\end{figure}

\begin{figure}[b]
\includegraphics[scale=0.455]{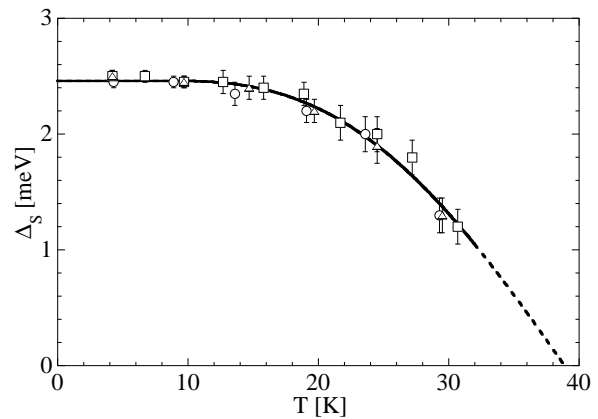}
\vskip-0.1in\caption{\label{dt}{Temperature evolution
$\Delta_S(T)$ from BCS fits to three independent sets of data
\cite{hSchmidt02}.}}
\end{figure}

\section{Temperature Dependence of $\Delta_S$}

The first unexpected result from tunneling spectroscopy on MgB$_2$
was the observation of a small energy gap, $\Delta_S=2.0$~meV
\cite{gRubio-Bollinger01}, much smaller than the weak coupling
limit would allow for $T_c=39$~K. The simplest scenario to result
in a reduced gap value is a surface layer of reduced
superconductivity with a gap anywhere in between the bulk value
and zero. This scenario can positively be ruled out by studying
the temperature dependence of the small gap, since such a reduced
gap should still scale with the local, reduced $T_c$, i.e.\ it
should obey ${2\Delta_0\over k_BT_c}\ge3.52$ \footnote{Enhanced
gap values are more easily understood as due to strong
electron--phonon coupling.}. This is illustrated in Fig.\
\ref{DsT}, where schematically the temperature dependence of an
intrinsic small gap is compared with the expectation for
$\Delta_S$ in a coupled two--band model.

For this reason, $\Delta_S(T)$ was measured in a number of
tunneling studies, and it was concluded to close at or near the
bulk $T_c$ of 39~K. However, the unusual combination of a small
gap and high temperatures results in significant smearing in SIN
junctions that renders the determination of $\Delta_S(T)$ more
difficult. In cases where two gaps are observed simultaneously the
exact analysis of high--temperature data is further complicated.
SIS junctions promise to provide a solution to this problem, as
the sharpness and position of the coherence peaks are insensitive
to thermal smearing.

\begin{figure}
\includegraphics[scale=0.45]{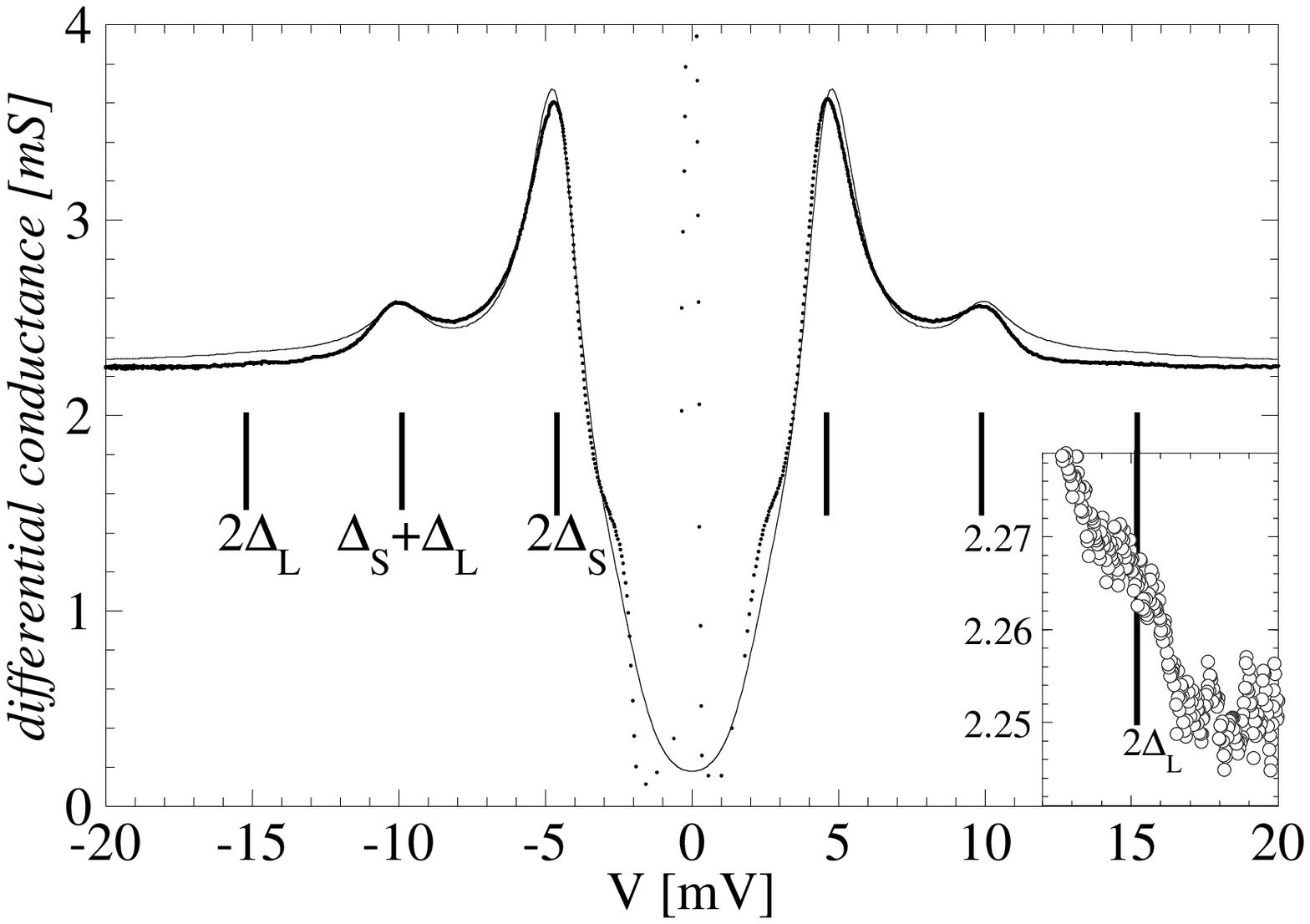}
\vskip-0.1in\caption{\label{doublepeak}{Conductance spectrum
showing direct current contribution from the 2D band along with a
fit to an SIS model using a weighted sum of two BCS DOSs. Inset
shows high bias on an expanded y--scale. The sharp zero--bias peak
reflects a Josephson current in this low--resistance junction.}}
\end{figure}

The evolution of SIS conductance spectra with temperature (Fig.\
\ref{temp}d) also allows a clear identification of a junction as
being SIS type, whereas the low--temperature data might be
inconclusive. To illustrate this, Fig.\ \ref{sinsis} shows a
junction that originally was interpreted in terms of SIN tunneling
using $\Delta=4.1$~meV and no smearing, but no temperature
dependent data was taken \cite{hSchmidt01,hSchmidt01b}. This fit
is compared to an SIS model using $\Delta=2.3$~meV and
$\Gamma=0.4$~meV, parameters that are now established to be
typical for MgB$_2$. Based only on the low--temperature data, a
conclusive decision is difficult, as both fits, though different
in detail, capture the major features of the data (with the
exception of the zero--bias peak \footnote{This peak may be due to
Josephson coupling in this medium--resistance junction
($R_N=1.2$~k$\Omega$). No such structures are seen in
high--resistance junctions (see e.g. Fig.\ \ref{temp-data},
$R_N=17.1$~k$\Omega$), whereas a sharp, hysteretic Josephson
current can be observed in low--resistance junctions (see e.g.
Fig.\ \ref{doublepeak}, $R_N=444$ $\Omega$).}).

Figure \ref{temp-data} finally shows the temperature dependence
for such a junction. Even at high temperature, a clearly developed
gap structure is visible in the raw data, which together with the
obvious closing of the peak position along with the closing of the
gap is conclusive evidence of the SIS nature of this junction. To
further analyze such data, a standard BCS SIS model was employed,
where $\Gamma$ was adjusted to best fit the lowest temperature
data and then held constant with increasing temperature, leaving
$\Delta_S$ the only parameter used to adjust the fits to higher
temperature data. These fits closely reproduce the evolution of
(i) the peak height, (ii) the zero bias conductance peak, and
(iii) the filling of the gap caused by thermally activated
quasiparticles. Thus all important features connected to the
coherence peaks are captured by this model, and we may safely use
it to determine $\Delta_S(T)$ from this data with high precision.

The results of such fits, performed on three different junctions,
are shown in Fig.\ \ref{dt}. The small error bars represent the
uncertainty in fitting the data, and there is excellent
reproducibility among the three sets of data. This evolution of
$\Delta_S$ closely resembles the theoretical prediction (see Fig.\
\ref{DsT}) for a two--band model, and is incompatible with this
gap being an isolated order parameter, as $2\Delta_0\over k_BT_c$
would be significantly lower than the weak--coupling limit. This
temperature dependence thus gives direct evidence for the presence
of a second, larger order parameter, although typically no
contribution from the second, 2D band is observed in SIS data.

\section{Spectroscopy of $\Delta_L$}

\begin{figure}
\includegraphics[scale=0.45]{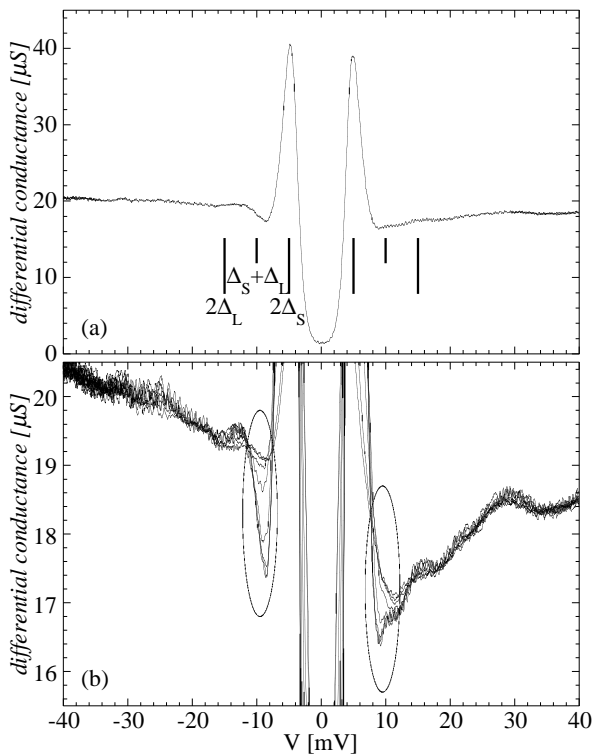}
\vskip-0.1in\caption{\label{dipmag}{(a) Low temperature spectrum
showing a clearly developed dip structure near
$\Delta_S+\Delta_L$. (b) Evolution of the background with
increasing field ($0-1.4$~T with a 0.2~T increment). The encircled
$\Delta_S+\Delta_L$--structure vanishes with increasing field.}}
\end{figure}

Such current contributions from the 2D band are expected to show
up at $eV=\Delta_L$ in SIN junctions and at $\Delta_S+\Delta_L$
and $2\Delta_L$ in SIS junctions. However, it is necessary to very
carefully analyze whatever extra features are observed. Both the
spectra in Fig.\ \ref{doublepeak} and Fig.\ \ref{dipmag}a,
respectively, show additional features at bias close to
$\Delta_S+\Delta_L$, yet the shape of these features is entirely
different. Whereas in the first case a distinct peak is observed,
the second spectra displays a dip at about the same position.

The extra peak in Fig.\ \ref{doublepeak} can easily be attributed
to direct tunneling contributions from the 2D band. The relative
weight of such contributions depends on the density of charge
carriers in the respective bands (the 3D band accounts for 58\% of
the total DOS \cite{ayLiu01}), and the orientation of the
junction. Since the small gap is hosted in a 3D band, it is
expected to contribute to junctions of any given orientation,
whereas the large gap requires the junction to be aligned with the
2D bands which generally will not be the case. We believe, that
the latter condition is crucial for the observed dominance of the
small gap.

Figure \ref{doublepeak} gives an example for the rare case that
such alignment is achieved in SIS break junctions. Two clearly
developed peaks are visible in the spectrum, that can be ascribed
to 3D--3D and 3D--2D tunneling. Note again the qualitative
difference between this second peak and the
$\Delta_S+\Delta_L$--dip feature that will be discussed below.

To fit this data, we use BCS DOSs for both bands, adjusting the
gap magnitudes to fit the well developed coherence peaks and
choosing the mixing ratios in both electrodes to reproduce the
observed peak heights. Such a model yields $\Delta_S=2.3$~meV and
$\Delta_L=7.6$~meV, consistent with our previous data (in the case
of $\Delta_S$) as well as with other experimental reports.

The fit shown in Fig.\ \ref{doublepeak} uses weights of 2 and 4\%,
respectively, for the large gap contribution in the two electrodes
\footnote{Note, that assuming the same $\Gamma$ for both bands the
$\Delta_S+\Delta_L$ peak is intrinsically sharper than the
$2\Delta_S$ peak and therefore the relative peak heights do not
directly reflect the $\sim94$\% contribution from 3D--3D and
$\sim6$\% from 3D--2D tunneling.}. While the gap magnitudes (which
contain the important, intrinsic information) are well--defined
from the data, this is not the case for these mixing ratios, as
there is no reason to assume them to be equivalent in both
electrodes. The presence of the second peak gives evidence for
contributions to the total current from the 2D band in at least
one electrode, however, it does not give evidence as to how this
2D contribution is distributed over both electrodes. In principle,
only one electrode needs to show alignment with the 2D band to
produce a second peak at $\Delta_S+\Delta_L$. The decisive
information would be the presence of a third peak at $2\Delta_L$,
however, its amplitude is expected to be negligible unless the the
2D band contributions are significant \emph{in both electrodes}
\footnote{The contributions to the effective total DOSs are
$c_{L1}$, $c_{L2}$ from the large and $c_{S1}=1-c_{L1}$,
$c_{S2}=1-c_{L2}$ from the small--gapped band in the 1$^{\rm st}$
and 2$^{\rm nd}$ electrode, respectively. Therefore, the
contribution to the conductance spectrum is $c_{S1}c_{S2}$ for the
$2\Delta_S$ peak, $c_{S1}c_{L2}+c_{L1}c_{S2}$ for the
$\Delta_S+\Delta_L$ peak, and finally $c_{L1}c_{L2}$ for the
$2\Delta_L$ peak. Thus a sizable effect from this high--bias peak
is observed only if the product of both 2D contributions is
sizable.}. We checked our data for such a third peak and find a
\emph{very} weak shoulder at the corresponding position that is
consistent with 0.1\% contribution to the total spectrum at most
(a blow--up of the high--bias data is given as an inset to Fig.\
\ref{doublepeak}). This is consistent with the weights of 2 and
4\% chosen for the fit (that yield 0.08\% contribution). Anyway,
the mixing ratio is determined by the relative orientation of the
grains and does not contain any intrinsic information. We
therefore accept the aforementioned consistency as sufficient.

\section{Two--Band Fitting of SIS Spectra}

Most tunneling work that determined $\Delta_L$, or at least found
evidence for its presence, did so by studying direct tunneling
contributions from the 2D band in SIN junctions (double--peak
structures). In the previous section we illustrated how such
contributions do, even though rarely, influence SIS junctions.
However, the more frequently seen extra feature in SIS junctions
is entirely different in shape and origin. Figure \ref{dipmag}
gives an example for a well--developed \emph{dip} near
$\Delta_S+\Delta_L$, in contrast to the \emph{peak} at this
position that we discussed before. This feature is commonly seen
in our SIS junctions, and can certainly not be explained by any
parallel tunneling channel, as such extra channels should
exclusively yield additional conductance.

Comparison of SIS fits and the actual data (see e.g.\ Fig.\
\ref{temp-data}) reveals that (with increasing bias) the data
first exceeds the BCS expectation (shoulder) and then
distinctively drops below it (dip) \footnote{In our best
junctions, the conductance indeed even drops below unity (see e.g.
Fig.\ \ref{dipmag}), however, in most junctions it {\lq\lq}at
least{\rq\rq} drops below the BCS line.}, a behavior reminiscent
of strong--coupling effects, where spectral weight is
redistributed in a similar fashion. However, at this energy there
are very few phonons available and the electron--phonon coupling
to the 3D band at any rate is expected to be weak. To establish
this feature to be intimately related to superconductivity, we
studied the whole background structure in more detail.

The bottom panel of Fig.\ \ref{dipmag} shows a magnification of
the conductance background in the junction shown on top,
displaying a variety of small structures, where the
$\Delta_S+\Delta_L$--feature by far is the most pronounced one.
With increasing magnetic field, the superconducting gap structure
weakens \footnote{A detailed study of the field dependence in
MgB$_2$ break junctions is in progress and will be reported
elsewhere.} and along with it, the $\Delta_S+\Delta_L$--feature
vanishes. At the same time, the other background structure remains
unchanged, proving the high stability of the junction and that the
$\Delta_S+\Delta_L$--dip is closely connected to
superconductivity.

\begin{figure}
\vskip-0.46in
\includegraphics[scale=0.45]{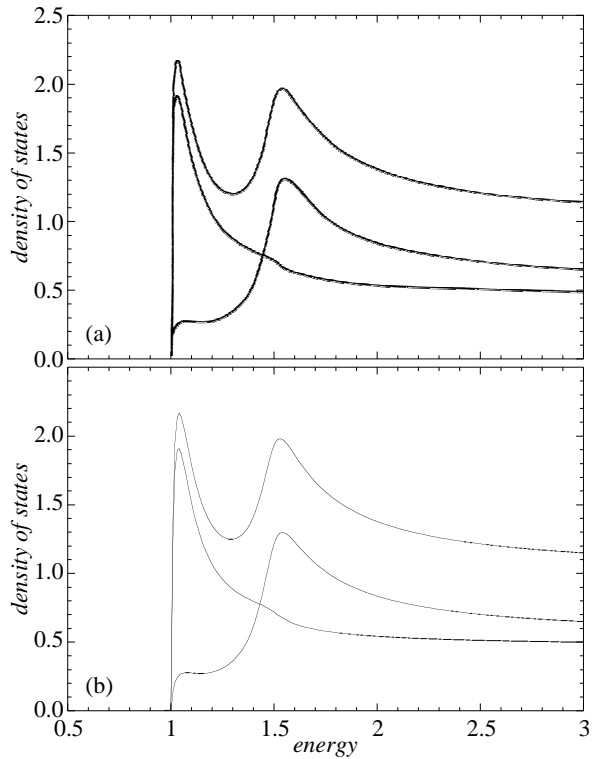}
\vskip-0.1in \caption{\label{schopohl}{(a) Partial and total DOS
in a full two--band model (reprinted from Ref.\ \cite{schopohl},
Fig.\ 3, copyright 1977, with permission from Elsevier Science).
(b) Equivalent DOS calculated using McMillan's tunneling model of
the proximity effect.}}
\end{figure}

We believe, that this extra feature is a signature of
quasiparticle transfer between the 2D and 3D bands. To explore
this, we need to abandon the simple BCS model used before and
employ a slightly more elaborate DOS, taking into account the
interband quasiparticle interaction. The coupling between both
bands in principle may be mediated by both pair and quasiparticle
transfer, however, only the quasiparticle contribution yields
corrections to the spectral shape of both energy gaps. Since the
original treatment of two--band superconductivity \cite{hSuhl59}
only considers pair transfer, it does not suffice for our purpose.
Quasiparticle transfer was included into a later theoretical
two--band model \cite{schopohl}, resulting in extra structure in
each DOS at the position of the respective other gap (see Fig.\
\ref{schopohl}). To keep modeling mathematically simple, we use an
observation pointed out in Ref.\ \cite{noce}, viz.\ that the DOS
in a pure BCS two--band model \cite{hSuhl59} and the DOS in
McMillan's tunneling model of the proximity effect
\cite{wlMcMillan68} are different only in that the latter includes
both pair and quasiparticle contribution on an equal footing. This
means, that the McMillan model may be used to generate the DOS in
a two--band model. The model requires the solution of two
simultaneous equations for the respective gap functions
\cite{wlMcMillan68}:
\begin{equation}
\Delta_{1}(E)={\Delta_{1}^{\rm ph}+ {\Gamma_{1}\Delta_{2}(E)/
\sqrt{\Delta_{2}^{2}(E)-(E-i\Gamma^{*}_{2})^{2}}}\over
1+{\Gamma_{1}/ \sqrt{\Delta_{2}^{2}(E)-(E-i\Gamma^{*}_{2})^{2}}}},
\end{equation}
\begin{equation}
\Delta_{2}(E)={\Delta_{2}^{\rm ph}+ {\Gamma_{2}\Delta_{1}(E)/
\sqrt{\Delta_{1}^{2}(E)-(E-i\Gamma^{*}_{1})^{2}}}\over
1+{\Gamma_{2}/ \sqrt{\Delta_{1}^{2}(E)-(E-i\Gamma^{*}_{1})^{2}}}},
\end{equation}
were $\Delta_{1,2}^{\rm ph}$ are the intrinsic pairing amplitudes
in both bands, $\Gamma_{1,2}$ are scattering rates related
inversely to the times spent in each band prior to scattering to
the other, and $\Gamma_{1,2}^*$ are smearing parameters in both
bands which were added to account for lifetime effects. These gap
functions are then used to create the DOS from the standard BCS
expression eq.\ (\ref{DOS}). SIS spectra are created using the
usual convolution eq.\ (\ref{SIS}).

To further verify the validity of this approach, we reproduced the
DOS calculated from a full two--band model \cite{schopohl} using
McMillan's model \cite{wlMcMillan68}. Fig.\ \ref{schopohl}
compares both results and an almost perfect agreement is found
over the entire energy range. We therefore may safely use this
model to fit our results and investigate the effects of interband
quasiparticle transfer on the DOS.

\begin{figure}
\includegraphics[scale=0.425]{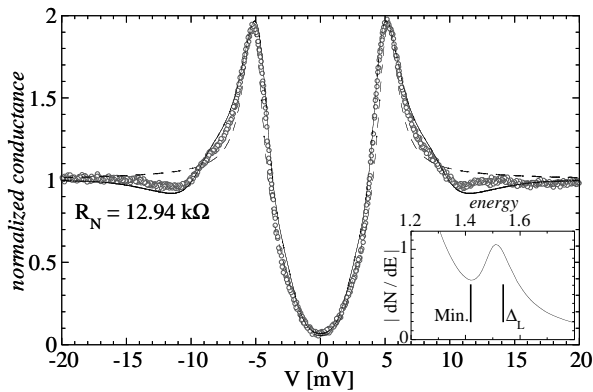}
\vskip-0.1in\caption{\label{prl-fig4}{Low temperature spectrum
showing direct tunneling exclusively from the 3D band, and an
additional dip feature near $\Delta_S+\Delta_L$. Thin dashed line
is a fit to a (one--band) BCS model, solid line to a two--band
model (McMillan's equations) \cite{hSchmidt02}. The parameters
used are $\Delta_1^{\rm ph}=0$, $\Delta_2^{\rm ph}=7.2$~meV,
$\Gamma_1=4.0$~meV, $\Gamma_2=1.0$~meV, $\Gamma_1^*=0.5$~meV, and
$\Gamma_2^*=1.0$~meV. See text for details. Inset: Derivative of
the small--gapped DOS shown in Fig.\ \ref{schopohl}b.}}
\end{figure}

Figure \ref{prl-fig4} gives an example for a typical
low--temperature SIS conductance spectrum, along with a fit to the
standard--BCS model used before (dashed line) and a fit using
McMillan's model (solid line). Whereas the BCS fit fails to
reproduce the higher bias features, McMillan's model results in a
close fit over the entire voltage range, \emph{including} the dip
near $\Delta_S+\Delta_L$ \footnote{Moca \emph{et al.} developed a
full two--band treatment of MgB$_2$ and achieved an even better
agreement with the very same data \cite{Moca-unpublished}.
However, the point of our model is to qualitatively demonstrate
the necessity to include quasiparticle transfer in any theoretical
model.}. To qualitatively understand the origin of this
dip--feature, consider the explicit form of the derivative of eq.
(\ref{SIS}) at zero temperature:
\begin{equation}
{dI\over dV}=\sigma_N \int_{-eV}^0N(E){\partial
N(E+eV)\over\partial eV}dE+\sigma_NN(-eV)N(0).
\end{equation}
The second term yields only a small correction (and vanishes if
there are no states at zero energy), thus the major contribution
consists of a convolution of the DOS and its derivative. The inset
to Fig.\ \ref{prl-fig4} shows the derivative of the small--gapped
DOS from Fig.\ \ref{schopohl}, displaying a pronounced minimum
just to the left of $\Delta_L$ (reflecting the shoulder in the DOS
at the same position). Convoluting this with the DOS, that shows a
sharp onset at $\Delta_S$, yields a dip in the conductance
slightly shifted from $\Delta_S+\Delta_L$ towards lower energy, as
we indeed find in our data (see also Fig.\ \ref{dipmag}a).

The success of this model confirms that the
$\Delta_S+\Delta_L$--dip reflects the quasiparticle coupling in
between both bands. To our knowledge, this so far is the only
piece of experimental evidence from tunneling spectroscopy to not
only show two gaps, but to also give direct evidence for their
coupling from a detailed analysis of their spectral shape.

\section{Ruling out Proximity Effects}

However, the applicability of this model  \emph{a priori} does not
prove two--band superconductivity to be the origin of the
dip--feature, as McMillan's model originally was developed to
describe the proximity effect, which therefore needs to be
considered as a possible cause. In the proximity effect, a thin
surface layer of reduced (or no) intrinsic superconductivity is
influenced by an adjacent superconductor to show enhanced
superconductivity (or some supercondutivity at all). McMillan's
equations specifically describe the proximity effect in the
presence of an insulating barrier of thickness, $d_I$, in between
the bulk superconductor and a surface layer of thickness, $d_N$.

Allowing for such an unknown surface layer (which in the case of
MgB$_2$ may easily be envisioned to be e.g.\ MgO$_x$) the
observation of a small gap is conceivable \emph{without}
considering two--band superconductivity. Note, that even the
temperature dependence of $\Delta_S$, or more specifically its
closing at $T_c$, is not inconsistent with the behavior of a
proximity sandwich. In spatially extended junctions (like point
contacts or planar junctions) it is furthermore plausible to find
more than one gap, as the surface layer may vary over the junction
area (or simply not be present in some areas)---such spatial
variations in sample properties over the junction area have been
suggested to explain point contact data.

There are two strong arguments against this proximity scenario and
in favor of two--band superconductivity. Firstly, the unusual
reproducibility of the small gap's magnitude in a wide variety of
experiments using different samples and tunneling techniques.
Referring again to Tab.\ \ref{gaps}, it is obvious, that almost
all experiments would agree to $\Delta_S=2.0-2.8$~meV. To fully
appreciate this agreement, one has to keep in mind how sensitively
the induced gap would depend on the properties of an assumed
proximity sandwich. $d_N$ enters the induced gap with a mere power
law, however, $d_I$,
---representing a tunneling barrier---enters exponentially (for an
experimental verification of this aspect of McMillan's model, see
Ref.\ \cite{keGray72}). This means, that very slight changes in
the barrier thickness result in a variation of the induced gap
that may easily cover several orders of magnitude.

\begin{figure}
\includegraphics[scale=0.27]{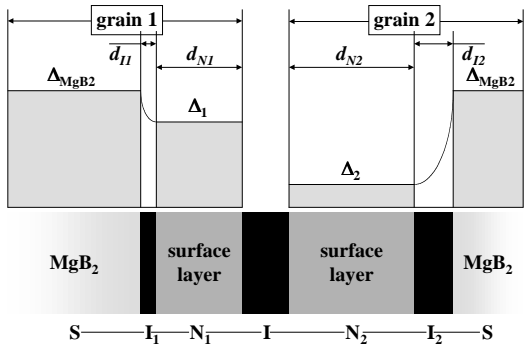}
\caption{\label{PIP}{Schematic representation of an asymmetric
SIS' junction caused by randomly, and non--uniformly formed
surface layers. Here, the surface layers are considered part of
the grains. See text for details.}}
\end{figure}

Again, our SIS geometry allows us to strengthen this point. Since
the junction becomes rather complex in this model, we refer to the
sketch in Fig.\ \ref{PIP}. In general, the SIS junction is now
represented by two proximity sandwiches separated by the main
barrier, I. Each sandwich in turn consists of the bulk
superconductor, S, a barrier, I$_{1,2}$, and a surface layer,
N$_{1,2}$. As the surface layers form randomly, there is no reason
to assume the I$_i$ and N$_i$ to be equivalent. Likewise, the
effective gap values entering the {\lq\lq}main{\rq\rq} SIS
junction may be different, resulting in an \emph{asymmetric} SIS'
junction.

Such junctions are known to show a unique spectral shape,
including a so--called {\lq\lq}difference peak{\rq\rq} at
$\Delta_1-\Delta_2$ \footnote{This peak is equivalent in origin to
the thermally activated zero--bias peak in symmetric junctions.}.
\emph{Vice versa}, the absence of this feature (as seen in Fig.\
\ref{temp-data}) gives direct evidence of the symmetry of the
junction. Testing for very small asymmetries requires a detailed
analysis of the in--gap spectral shape, since the difference peak
(or zero--bias peak in symmetric junctions) is subject to thermal
broadening---in stark contrast to the coherence peak (see Fig.\
\ref{temp}d). Therefore, the difference peak may be veiled and
show up only as a characteristic change of the in--gap spectral
shape. Analyzing simulated SIS' spectra allowing for increasing
difference in gap values, we find that 20\% difference is about
the maximum that might still be considered consistent with our
data. Using the numbers from Ref.\ \cite{keGray72}, we estimate
that this translates into a reproducibility of barrier thickness,
I$_{1,2}$, on the order of $\pm0.5$~\AA. Considering the initial
assumption that such proximity sandwiches need to form randomly,
and independently, such a reproducibility certainly proves this
consideration to be highly unlikely.

A second argument against the proximity effect is the observation
of both gaps simultaneously at a \emph{single point} in STM
studies, which is very hard to reconcile with the assumption of
variable surface layers. Although in principle any proximity
induced, small gap structure is expected to show an additional
feature at the intrinsic, large gap energy, this feature is
expected to be much weaker than the main, induced gap structure.
It certainly is inconsistent with two peaks of equal magnitude, or
even the high--bias peak being more pronounced than the low--bias
peak (see e.g.\ data in Ref.\ \cite{Iavarone-unpublished}).

The appearance of two distinct gap values, the temperature
dependence of the small gap, and the appearance of additional
spectral features are therefore not only more easily understood in
the framework of two--band superconductivity, but---beyond
that---not readily consistent with the proximity effect.

\section{Conclusion}

We have reproducibly observed a small gap, $\Delta_S=2.5$~meV, in
break--junction tunneling on MgB$_2$ and traced it to high
temperatures, where it closes near the bulk $T_c$. Only in rare
cases, we also observe direct tunneling contributions showing a
large gap, $\Delta_L=7.6$~meV. These findings give evidence for,
and are interpreted in terms of a two--band model. A commonly
observed dip feature at $\Delta_S+\Delta_L$ is analyzed using a
specific two--band model. We argued, that this feature gives
evidence for quasiparticle contributions to interband coupling,
and furthermore showed it to be inconsistent with the result of
proximity effects. Taken together with tunneling data from the
literature, this gives very convincing evidence that MgB$_2$ is
one of the very rare materials showing two--band
superconductivity.

\section*{Acknowledgements}

We are indebted to B.\ Jank\'{o} and C.\ P.\ Moca for valuable
discussions and sharing unpublished results from their two--band
calculations. We want to thank K.\ Scharnberg for allowing us to
use Fig.\ \ref{schopohl}a. We further want to thank M.\ A.\
Belogolovskii, G.\ Burnell, D.\ J.\ Kang, J.--I. Kye, D.\
Mijatovic, A.\ Saito, and K.\ Ueda for sending their manuscripts
prior to publication. This research is supported by the U.S.\
Department of Energy, Basic Energy Sciences---Materials Sciences,
under contract \#~W--31--109--ENG--38.


\begin{thebibliography}{00}
\bibitem{jNagamatsu01} J.\ Nagamatsu \emph{et al.}, Nature
\textbf{410} (2001) 63.
\bibitem{ayLiu01} A.\ Y.\ Liu \emph{et al.}, Phys.\
Rev.\ Lett.\ \textbf{87} (2001) 087005.
\bibitem{iGiaver60} I.\ Giaever, Phys.\ Rev.\ Lett.\ \textbf{5}
(1960) 147.
\bibitem{rcDynes78} R.\ C.\ Dynes, V.\ Narayanamurti and J.\ P.\ Garno,
Phys.\ Rev.\ Lett.\ \textbf{41} (1978) 1509.
\bibitem{afAndreev64} A.\ F.\ Andreev,
Zh.\ \'{E}ksp.\ Teor.\ Fiz.\ \textbf{46} (1964) 1823 [Sov.\ Phys.\
JETP \textbf{19} (1964) 1228].
\bibitem{geBlonder82} G.\ E.\ Blonder, M.\ Tinkham and T.\ M.\ Klapwijk, Phys.\
Rev.\ B \textbf{25} (1982) 4515.
\bibitem{aPlecenik94} A.\ Plecen\'{i}k \emph{et al.}, Phys.\ Rev.\
B \textbf{49} (1994) 10016.
\bibitem{gRubio-Bollinger01} G.\ Rubio--Bollinger \emph{et al.},
Phys.\ Rev.\ Lett.\ \textbf{86} (2001) 5582.
\bibitem{hSuderow02} H.\ Suderow \emph{et al.}, Physica C
\textbf{369} (2002) 106.
\bibitem{aSharoni01a} A.\ Sharoni \emph{et al.}, J.\ Phys.:\
Cond.\ Mat.\ \textbf{13} (2001) L503.
\bibitem{aSharoni01b} A.\ Sharoni \emph{et al.}, Phys.\ Rev.\ B
\textbf{63} (2001) 220508.
\bibitem{gKarapetrov01} G.\ Karapetrov \emph{et al.}, Phys.\
Rev.\ Lett.\ \textbf{86} (2001) 4374.
\bibitem{gKarapetrov01b} G.\ Karapetrov \emph{et al.}, in \emph{Studies of High Temperature
Superconductors}, ed.\ A.\ V.\ Narlikar, (Nova Sci. Publ., New
York, 2002), vol.\ 38, pp.\ 221.
\bibitem{Iavarone-unpublished} M.\ Iavarone \emph{et al.}, Phys.\ Rev.\ Lett.\
\textbf{89} (2002) 187002.
\bibitem{Olsson-unpublished} R.\ J.\ Olsson \emph{et al.},
unpublished (cond-mat/0201022).
\bibitem{fGiubileo01} F.\ Giubileo \emph{et al.}, Phys.\ Rev.\
Lett.\ \textbf{87} (2001) 177008.
\bibitem{fGiubileo02a} F.\ Giubileo \emph{et al.}, Europhys.\
Lett.\ \textbf{58} (2002) 764.
\bibitem{fGiubileo02b} F.\ Giubileo \emph{et al.}, Int.\ J.\ Mod.\
Phys.\ B \textbf{16} (2002) 1577.
\bibitem{Karpinski-unpublished} J.\ Karpinski \emph{et al.},
unpublished  (cond--mat/0207263).
\bibitem{Eskildsen-unpublished} M.\ R.\ Eskildsen \emph{et al.},
Phys.\ Rev.\ Lett.\ \textbf{89} (2002) 187003.
\bibitem{pSeneor01} P.\ Seneor \emph{et al.}, Phys.\ Rev.\ B
\textbf{65} (2001) 012505.
\bibitem{aPlecenik02} A.\ Plecenik \emph{et al.}, Physica C
\textbf{368} (2001) 251.
\bibitem{fLaube01} F.\ Laube \emph{et al.}, Europhys.\ Lett.\
\textbf{56} (2001) 296.
\bibitem{Dyachenko-unpublished} A.\ I.\ D'yachenko \emph{et al.}, unpublished (cond-mat/0201200).
\bibitem{zzLi01} Z.\ Z.\ Li \emph{et al.}, Supercond.\ Sci.\
Technol.\ \textbf{14} (2001) 994.
\bibitem{zzLi02b} Z.--Z.\ Li \emph{et al.}, Phys.\ Rev.\ B
\textbf{66} (2002) 064513.
\bibitem{pSzabo01} P.\ Szab$\rm\acute{o}$ \emph{et al.}, Phys.\ Rev.\
Lett.\ \textbf{87} (2001) 137005.
\bibitem{ygNaidyuk02} Yu.\ G.\ Naidyuk \emph{et al.}, Pis.\ Zh.\ \'{E}ksp.\ Teor.\ Fiz.\
\textbf{75} (2002) 283 [JETP Lett.\ \textbf{75} (2002) 238].
\bibitem{Bobrov-unpublished} N.\ L.\ Bobrov \emph{et al.}, in \emph{New Trends in
Superconductivity}, ed. J.\ F.\ Annett and S.\ Kruchinin, (Kluwer
Academic Publ., Dodrecht, 2002), vol.\ 67, pp.\ 225.
\bibitem{Yanson-unpublished} I.\ K.\ Yanson \emph{et al.}, unpublished (cond--mat/0206170).
\bibitem{Lee-unpublished} S.\ Lee \emph{et al.}, Physica C
\textbf{377} (2002) 202.
\bibitem{yBugoslavsky02} Y.\ Bugoslavsky \emph{et al.}, Supercond.\ Sci.\
Technol.\ \textbf{15} (2002) 526.
\bibitem{aKohen01} A.\ Kohen \emph{et al.}, Phys.\ Rev.\ B
\textbf{64} (2001) 060506.
\bibitem{Belogolovskii-private} M.\ A.\ Belogolovskii \emph{et al.}, unpublished.
\bibitem{rsGonnelli02b} R.\ S.\ Gonnelli \emph{et al.}, J.\ Phys.\
Chem.\ Solids in press.
\bibitem{Gonnelli-unpublished} R.\ S.\ Gonnelli \emph{et al.},
unpublished (cond--mat/0208060).
\bibitem{rsGonnelli02a} R.\ S.\ Gonnelli \emph{et al.}, Int.\ J.\
Mod.\ Phys.\ B \textbf{16} (2002) 1553.
\bibitem{rsGonnelli01} R.\ S.\ Gonnelli \emph{et al.}, Phys.\
Rev.\ Lett.\ \textbf{87} (2001) 097001.
\bibitem{hSchmidt01} H.\ Schmidt \emph{et al.}, Phys.\ Rev.\ B
\textbf{63} (2001) 220504.
\bibitem{hSchmidt01b} H.\ Schmidt \emph{et al.}, in
\emph{Studies of High Temperature Superconductors}, ed.\ A.\ V.\
Narlikar, (Nova Sci Publ., New York, 2002), vol.\ 38, pp.\ 229.
\bibitem{hSchmidt02} H.\ Schmidt \emph{et al.}, Phys.\ Rev.\
Lett.\ \textbf{88} (2002) 127002.
\bibitem{hSchmidt-unpublished} present work.
\bibitem{Takasaki-unpublished} T.\ Takasaki \emph{et al.}, Physica
C \textbf{378--381} (2002) 229.
\bibitem{yZhang01} Y.\ Zhang \emph{et al.}, Appl.\ Phys.\ Lett.\
\textbf{79} (2001) 3995.
\bibitem{yXuan01} Y.\ Xuan \emph{et al.}, Chin.\ Phys.\ Lett.\
\textbf{18} (2001) 1254.
\bibitem{zzLi02} Z.\ Z.\ Li \emph{et al.}, Physica C \textbf{370}
(2002) 1.
\bibitem{Kye-unpublished} J.--I.\ Kye \emph{et al.}, unpublished. 
\bibitem{dMijatovic02} D.\ Mijatovic \emph{et al.}, Appl.\ Phys.\
Lett.\ \textbf{80} (2002) 2141.
\bibitem{Mijatovic-unpublished} D.\ Mijatovic \emph{et al.},
unpublished.
\bibitem{aBrinkman01b} A.\ Brinkman \emph{et al.}, Appl.\ Phys.\
Lett.\ \textbf{79} (2001) 2420.
\bibitem{gBurnell01} G.\ Burnell \emph{et al.}, Appl.\ Phys.\ Lett.\
\textbf{79} (2001) 3464.
\bibitem{gBurnell02} G.\ Burnell \emph{et al.}, Appl.\ Phys.\
Lett.\ \textbf{81} (2002) 102.
\bibitem{Burnell-unpublished} G.\ Burnell \emph{et al.},
unpublished. 
\bibitem{Kang-unpublished} D.--J.\ Kang \emph{et al.}, Appl.\ Phys.\ Lett.\ \textbf{81}
(2002) 3600.
\bibitem{Kang-unpublished2} D.--J.\ Kang \emph{et al.},
unpublished. 
\bibitem{dkAswal02} D.\ K.\ Aswal \emph{et al.}, Phys.\ Rev.\ B
\textbf{66} (2002) 012513.
\bibitem{mhBadr02} M.\ H.\ Badr \emph{et al.}, Phys.\ Rev.\ B \textbf{65} (2002) 184516.
\bibitem{gCarapella02} G.\ Carapella \emph{et al.}, Appl.\ Phys.\
Lett.\ \textbf{80} (2002) 2949.
\bibitem{Saito-unpublished} A.\ Saito \emph{et al.}, unpublished. 
\bibitem{Ueda-unpublished} K.\ Ueda and M.\ Naito, unpublished (cond--mat/0208571). 
\bibitem{hjChoi02a} H.\ J.\ Choi \emph{et al.}, Phys.\ Rev.\ B
\textbf{66} (2002) 020513(R).
\bibitem{hjChoi02b} H.\ J.\ Choi \emph{et al.}, Nature
\textbf{418} (2002) 758.
\bibitem{lOzyuzer98} L.\ Ozyuzer, J.\ F.\ Zasadzinski and K.\ E.\ Gray, Cryogenics
\textbf{38} (1998) 911.
\bibitem{hSuhl59} H.\ Suhl, B.\ T.\ Matthias and L.\ R.\ Walker,
Phys.\ Rev.\ Lett.\ \textbf{3} (1959) 552.
\bibitem{schopohl} N.\ Schopohl and K.\ Scharnberg, Solid State Commun. \textbf{22} (1977) 371.
\bibitem{noce} C.\ Noce and L.\ Maritato, Phys.\ Rev.\ B \textbf{40} (1989) 734.
\bibitem{wlMcMillan68} W.\ L.\ McMillan, Phys.\ Rev.\ \textbf{175} (1968)
537.
\bibitem{Moca-unpublished} C.\ P.\ Moca \emph{et al.}, unpublished (cond-mat/0210445).
\bibitem{keGray72} K.\ E.\ Gray, Phys.\ Rev.\ Lett.\ \textbf{28}
(1972) 959.
\end{thebibliography}
\end{document}